\documentclass[aip,apl,preprint]{revtex4-1}

\usepackage{bm}
\usepackage{graphicx}   
\usepackage[hang]{subfigure}

\begin{document}

\title {Josephson Tunnel Junctions in a Magnetic Field Gradient}
\thanks{Submitted to Appl. Phys. Letts.}

\author{R. Monaco}
\affiliation{Istituto di Cibernetica del CNR, Comprensorio Olivetti, I-80078, Pozzuoli, Italy and Dipartimento di Fisica, Universit$\grave{\rm a}$ di Salerno, 84081 Baronissi, Italy}\email
{roberto@sa.infn.it}

\author{J. Mygind}
\affiliation{DTU Physics, B309, Technical University of
Denmark, DK-2800 Lyngby, Denmark}
\author{V.\ P.\ Koshelets}
\affiliation{Kotel'nikov Institute of Radio Engineering and Electronics, Russian Academy of Science, Mokhovaya 11, Bldg 7, 125009, Moscow, Russia.}

\date{\today}
\begin{abstract}
We measured the magnetic field dependence of the critical current of high quality $Nb$-based planar Josephson tunnel junctions in the presence of a controllable non-uniform field distribution. We found skewed and slowly changing magnetic diffraction patterns quite dissimilar from the Fraunhofer-like ones typical of a homogeneous field. Our findings can be well interpreted in terms of recent theoretical predictions [R. Monaco, {\it J. Appl. Phys.}{\bf 108}, 033906 (2010)] for a uniform magnetic field gradient leading to Fresnel-like magnetic diffraction patterns. We also show that Fiske resonances can be suppressed by an asymmetric magnetic field profile.
\end{abstract}

\maketitle

It is commonly believed that a Faunnhofer-like magnetic diffraction pattern (MDP) characterized by its periodic and well-pronounced minima is the hallmark of the d.c. Josephson effect\cite{joseph,barone} in ideal tunnel junctions, weak links, Dayem bridges etc. However, this is only true if the applied magnetic field is uniform over the junction area. Experimentally this is rarely satisfied: in fact, the presence of the tip of a magnetic force microscope, of trapped Abrikosov vortices, of magnetic dots or strips, the self-field of nearby wires and/or electrodes, the mounting out of a solenoid axis etc., make the magnetic field distribution in the Josephson device environment non-uniform. Recently\cite{JAP10} the consequences of a uniform magnetic field {\it gradient} on the static properties of both electrically {\it short} and {\it long} planar Josephson tunnel junctions (JTJs) have been studied analytically and numerically, respectively. In both cases marked differences from the ideal text-book case\cite{barone} of perfectly homogeneous magnetic field were predicted: short junctions exhibit zeros-free Fresnel-like MDPs characterized by small-amplitude non-periodic damped oscillations slowly approaching a non-zero large-field asymptotic value and by symmetry with respect to current {\it or} field inversion. Upon increasing the junction normalized length $l$ the MDPs show a progressive break of these symmetries and only the symmetry with respect to the current {\it and} field inversion is retained. In this Letter, we report measurements of the magnetic field dependence of the critical current and of the amplitude of resonant cavity modes (Fiske steps) of high-quality window-type $Nb/Al-Al_{ox}/Nb$ JTJs in presence of a static non-uniform magnetic field. We believe that our finding can be very useful to correctly interpret those non-Fraunhofer-like MDPs which may erroneously be attributed to low sample quality; in addition we provide hints for a deeper understanding of both the static and dynamics properties of Josephson devices subjected to non-uniform magnetic fields.

\begin{figure}[t]
\centering
\subfigure[]{\includegraphics[width=6cm]{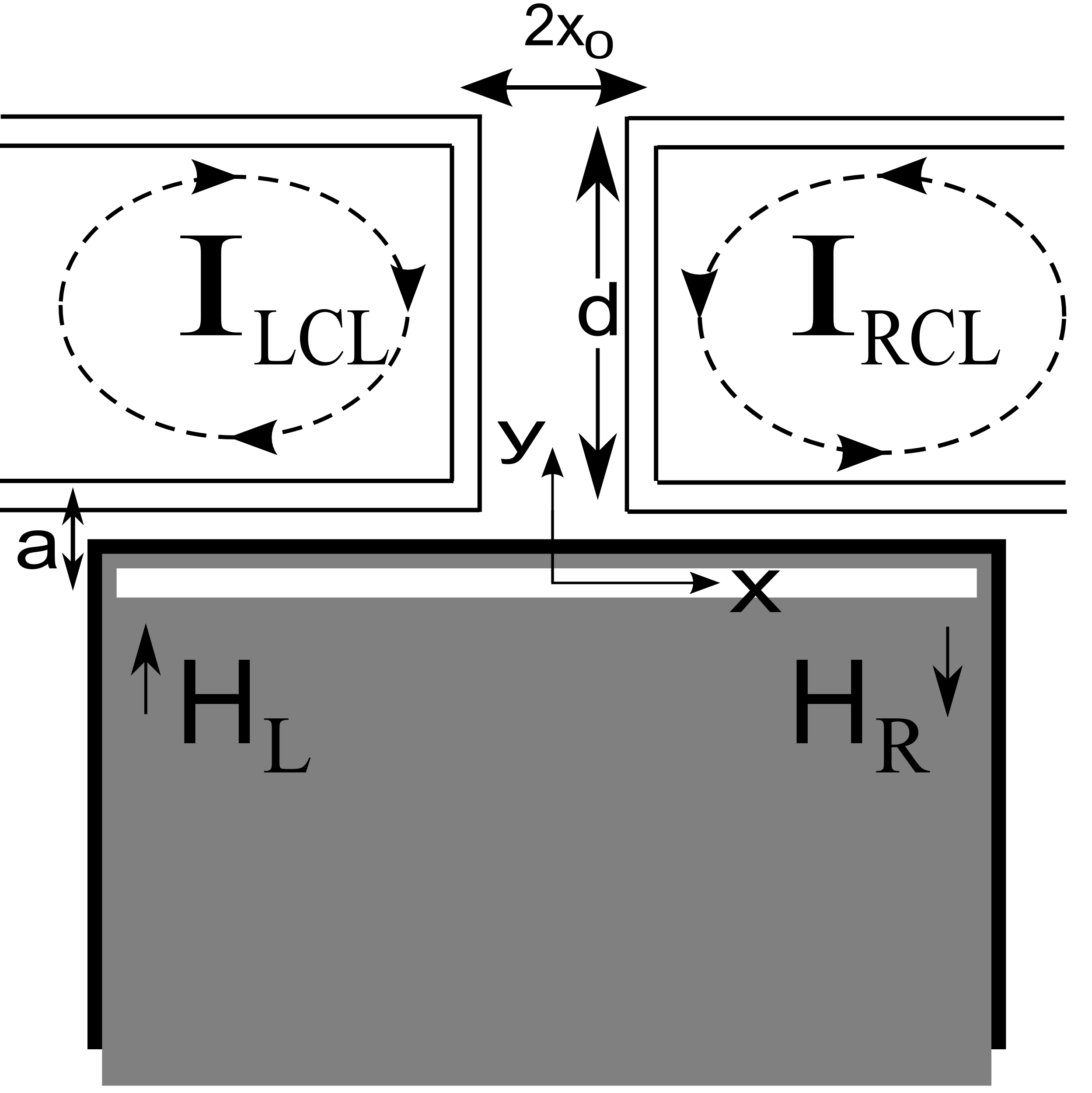}}
\subfigure[]{\includegraphics[width=8cm]{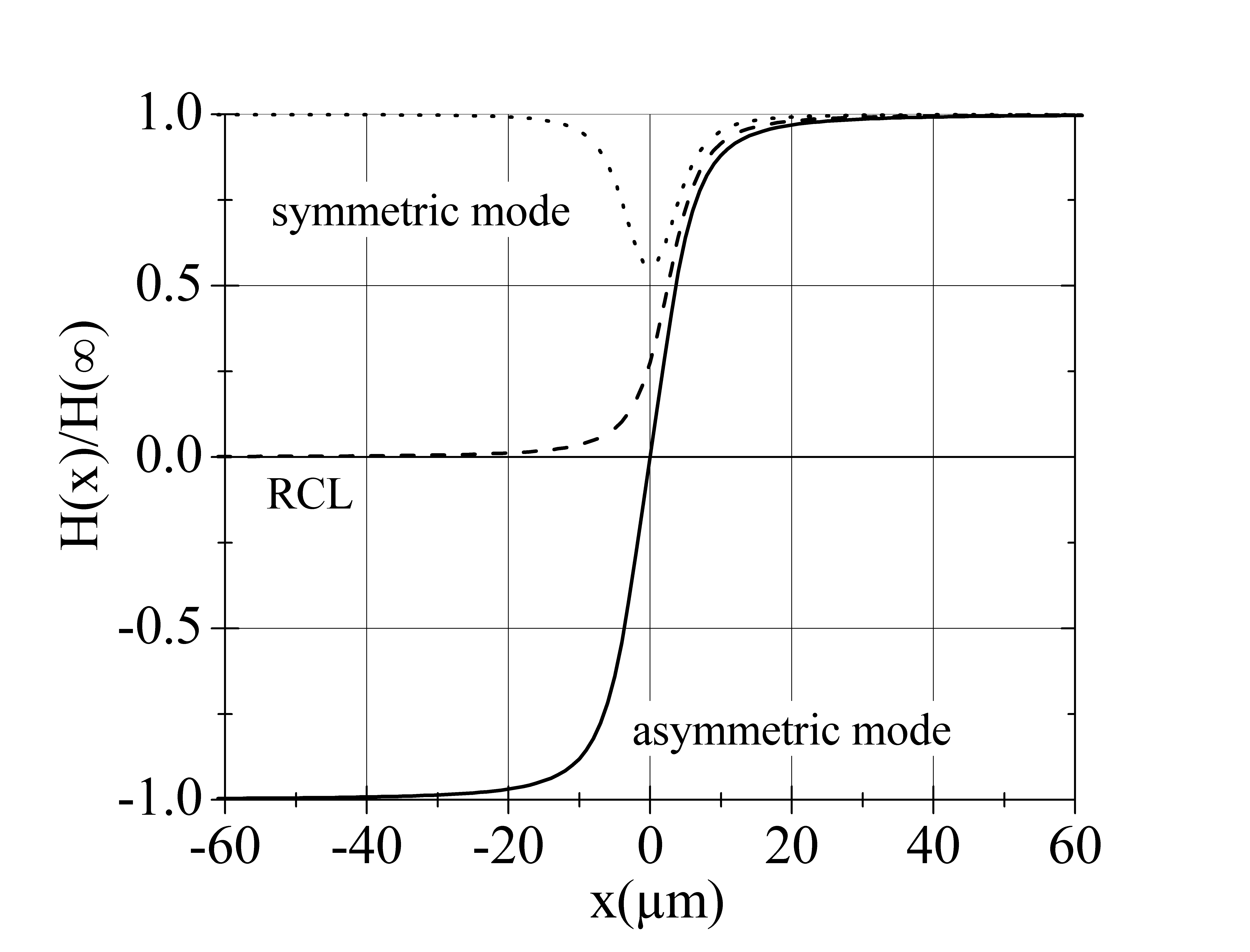}}
\caption{(a) Sketch (not in scale) of a $(2 \times 100) \mu m^2$  window-type Josephson tunnel junctions in the asymmetric magnetic field profile $H(x)$ generated by two properly biased independent control lines. The junction area is white, the base electrode black and the top electrode gray. In our design $a=2x_0=5 \mu m$ and $d=600\mu m$. (b) Computed magnetic field profiles along the junction length in the RCL, symmetric and asymmetric modes (dashed, solid and dotted line, respectively), for the layout depicted in (a).}
\end{figure}

Our design is sketched in Fig.1(a) together with the coordinate system used in this work; two independent $5 \mu m$ wide and $0.5 \mu m$ thick $Nb$ control lines run adjacent to the long dimension of an asymmetric $(2 \times 100) \mu m^2$ window-type JTJ, $2 \mu m$ away from the border of the junction base electrode. In the vicinity of the junction center they bent at right angle and run parallel to each other separated by a $2 \mu m$ gap. At a distance $d=600 \mu m$ from the first bending the lines bend back and run again in opposite directions. In case the left and right control line currents, $I_{LCL}$ and $I_{RCL}$ respectively, have the same amplitudes and the directions indicated by the left clockwise and right counterclockwise arrows in Fig.1(a), then an asymmetric field profile is realized; since the magnetic fields at the junction extremities have the same amplitude, but opposite directions, $H_L=-H_R$, we call {\it asymmetric} mode. In the opposite case, we end up with a symmetric profile field similar to the classical uniform field generated by a single control line running aside the whole junction length (or by a long solenoid); we will refer to it as the {\it symmetric} mode. By playing separately with $I_{LCL}$ and $I_{RCL}$, one can achieve any desired magnetic boundary condition. The largest achievable field values are set by the control line critical currents which depend on the temperature and the geometrical parameters. The control line technique has been used to produce local magnetic fields for digital applications of Josephson circuits since 1969\cite{matisoo}. It is useful to remind that the current $I_{CL}$ in each control line produces a magnetic field perpendicular to the junction plane which induces Meissner screening currents in the superconducting junction electrodes; these circulating currents, in turn, produce a magnetic field in the junction plane proportional to $I_{CL}$. The details of how a transverse field modulates the critical current of planar JTJs with different geometrical configurations can be found in Ref.6. Considering that the distance $d$ between the parallel arms of each control line is orders of magnitude larger than the separation, $a$, between the junction and the control lines, the magnetic field distribution $H(x,y=a)$ along the junction long dimension generated by each control line can be evaluated, to a very good approximation, by applying the Laplace equation to a semi-infinite wire carrying a current $\pm I_{CL}$, originating at $x=\pm x_0$ and running along the $x$ axis. For the right control line (RCL) operated alone we get $H_{RCL}(x,a) = H(\infty,a) [0.5 + 0.5 \sin \tan^{-1} (x-x_0)/a]$ with $H(\infty,a)= I_{CL}/(4\pi a)$, as shown by the dashed line in Fig.1(b) in which the solid and dotted lines show, respectively, the field profile for the asymmetric and symmetric modes. We observe that the largest field changes occur in a region in the middle of the junction whose size is determined by the distance $2x_0$ between the left and right control lines; in our case this separation is much smaller than the junction physical length. Quite similar field distributions were computed with 3D magneto-static Comsol Multiphysics simulations taking into account the correction to the free-space solution due to the presence of close superconducting electrodes.

\begin{figure}[t]
\includegraphics[width=8cm]{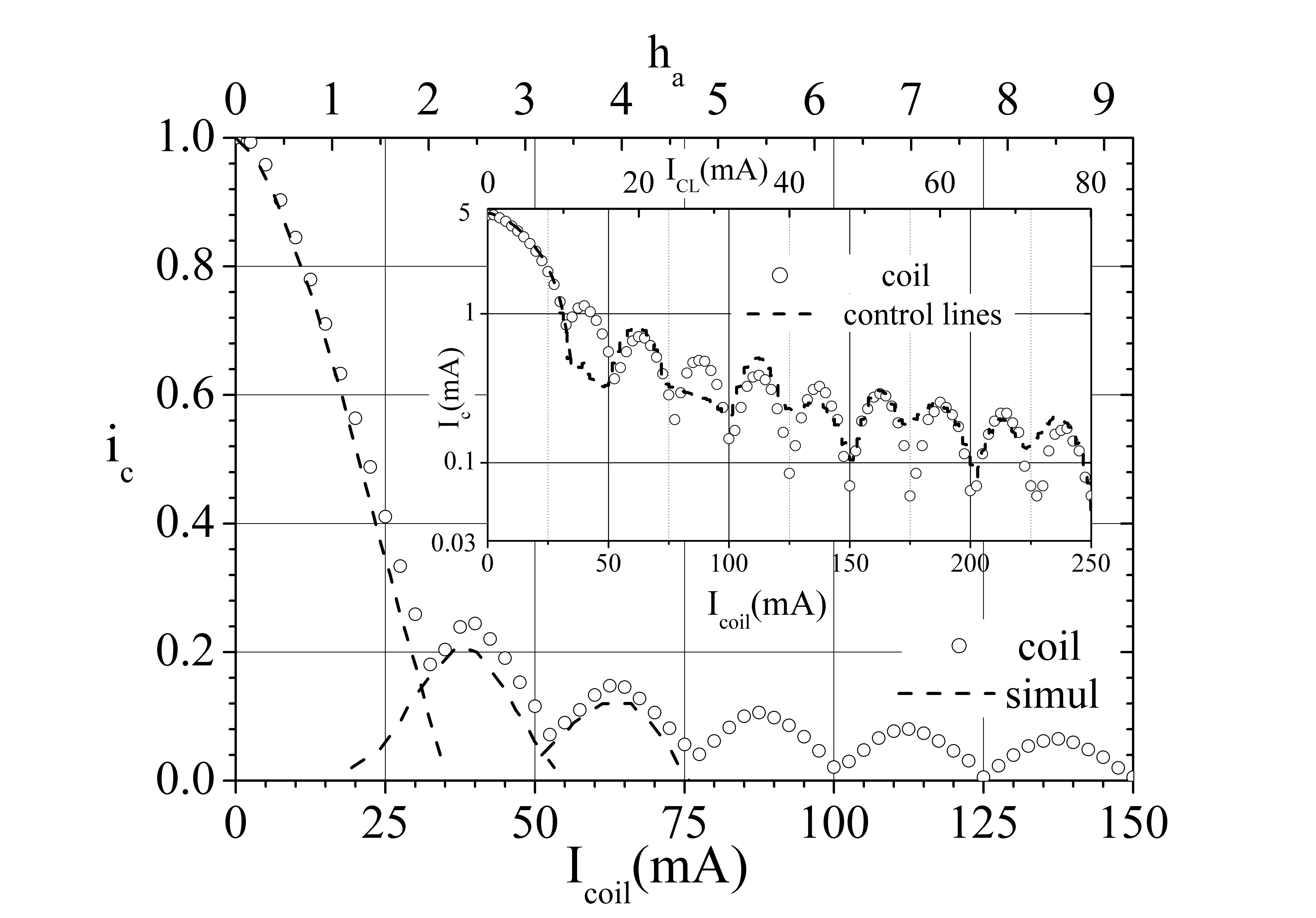}
\caption{Normalized MDP obtained in the uniform field generated by a long coil (open circles and lower horizontal scale) compared with the computed $i_c(h_a)$ for a one-dimensional overlap junction with normalized length $l=4$  in a uniform normalized field $h_a$ (dashed line and upper horizontal scale). The inset compares on a logarithmic scale the MDPs obtained in the uniform field of a solenoid (open circles and lower horizontal scale) and by the control lines in the symmetric mode (dotted line and upper horizontal scale). $T=4.2K$ and $I_c(0)=4.85mA$.}
\end{figure}

In the following, we will present the data of one representative sample out of few tested ones all having Josephson current density of $2.5 kA/cm^2$ at $T=4.2K$ and $(2 \times 100) \mu m^2$ barrier area. We underline that the overlap geometry is the only one for which is strictly proper to speak of zero-field behavior; for other geometries, current-bias self-induced fields are not negligible. Overlap-type JTJs with asymmetric electrode configuration have often been considered in the literature but, to our knowledge, only from a theoretical point of view; the two overlapping electrodes make the bias current distribution very uniform over the junction length and at the same time create a very large (asymmetric) idle region which {\it dresses} the junction and strongly influences its behavior with respect to that of bare junctions\cite{JAP7&8} (increased Josephson penetration depth and Swihart velocity). Preliminarily, we characterize the sample by measuring its MDP in the (conventional) uniform field of a long solenoid whose axis lies in the barrier plane and is perpendicular to the junction long dimension, i.e., along the $y$-direction of Fig.1(a). Such a test MDP is shown, in normalized units by the solid circles of Fig.2; $i_c=I_c/I_c(0)$ with $I_c(0)=4.85mA$ at $T=4.2K$. As the dotted line indicates, the experimental data are best fitted by the numerically computed in-plane MDP of an overlap junction when its normalized length $l=L/\lambda_j$ is set to $4$, indicating that the sample Josephson penetration depth is $\lambda_j \approx 25 \mu m$. (For a naked junction with no idle region the calculated $\lambda_j$ would be approximately equal to $7 \mu m$, with the $Nb$ London penetration set equal to $90nm$.) In the inset of Fig.2 we compare the same data with that obtained with the control lines operated in the symmetric mode (dotted line and upper horizontal scale). The vertical log-scale is chosen to emphasize the suppression of few pattern lobes. This is likely due to the non uniformity of the field profile in the symmetric mode - see the well in the dotted line of Fig.1(b). Incidentally we observe that, despite the boundary conditions are the same, $H_L=H_R$, the MDPs are quite different. We remark that all the MDPs in Fig.2 and its inset are symmetric with respect to inversion of either the junction bias current or the applied magnetic field.

\begin{figure}[ht]
\subfigure[ ]{\includegraphics[width=7cm]{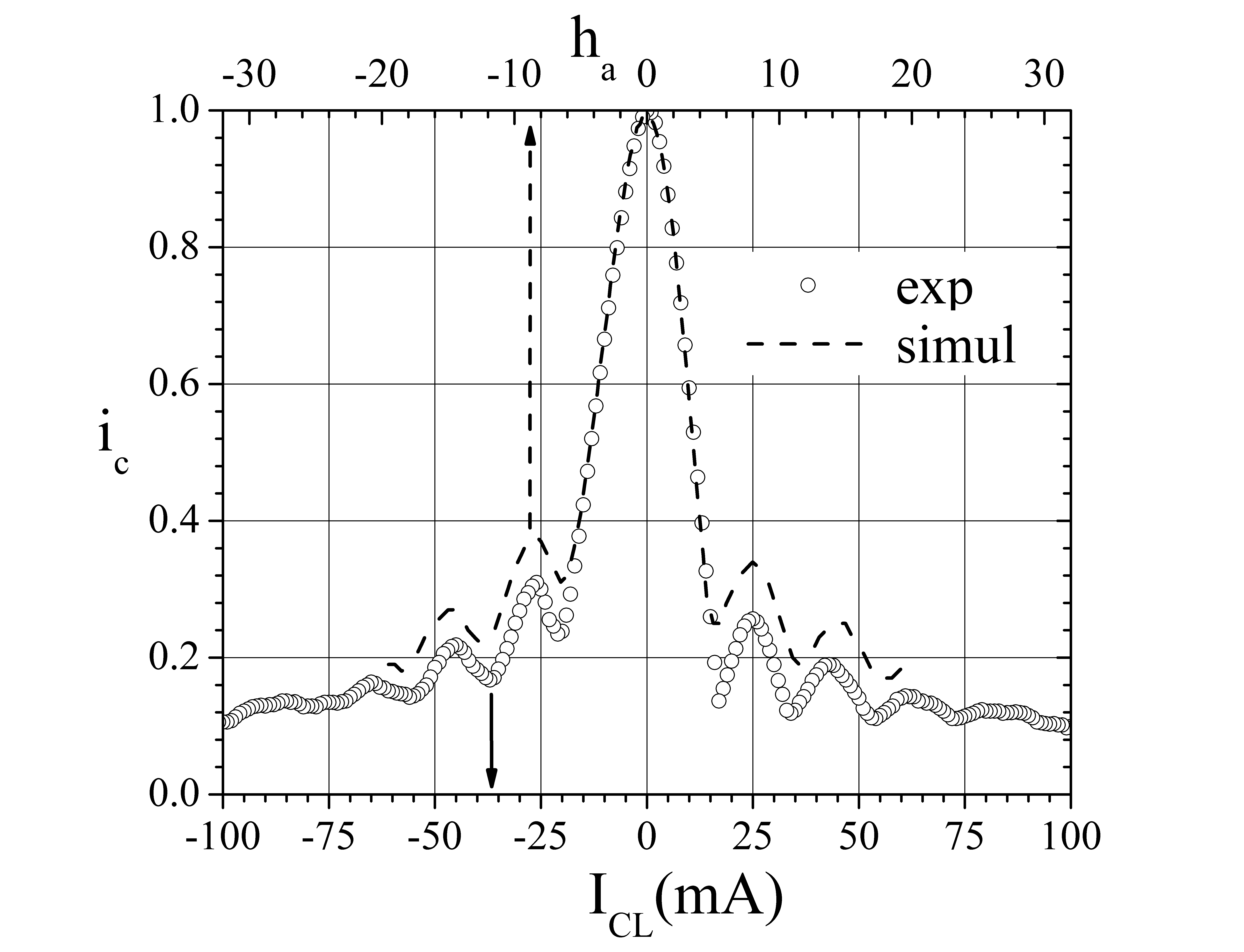}}
\subfigure[ ]{\includegraphics[width=7cm]{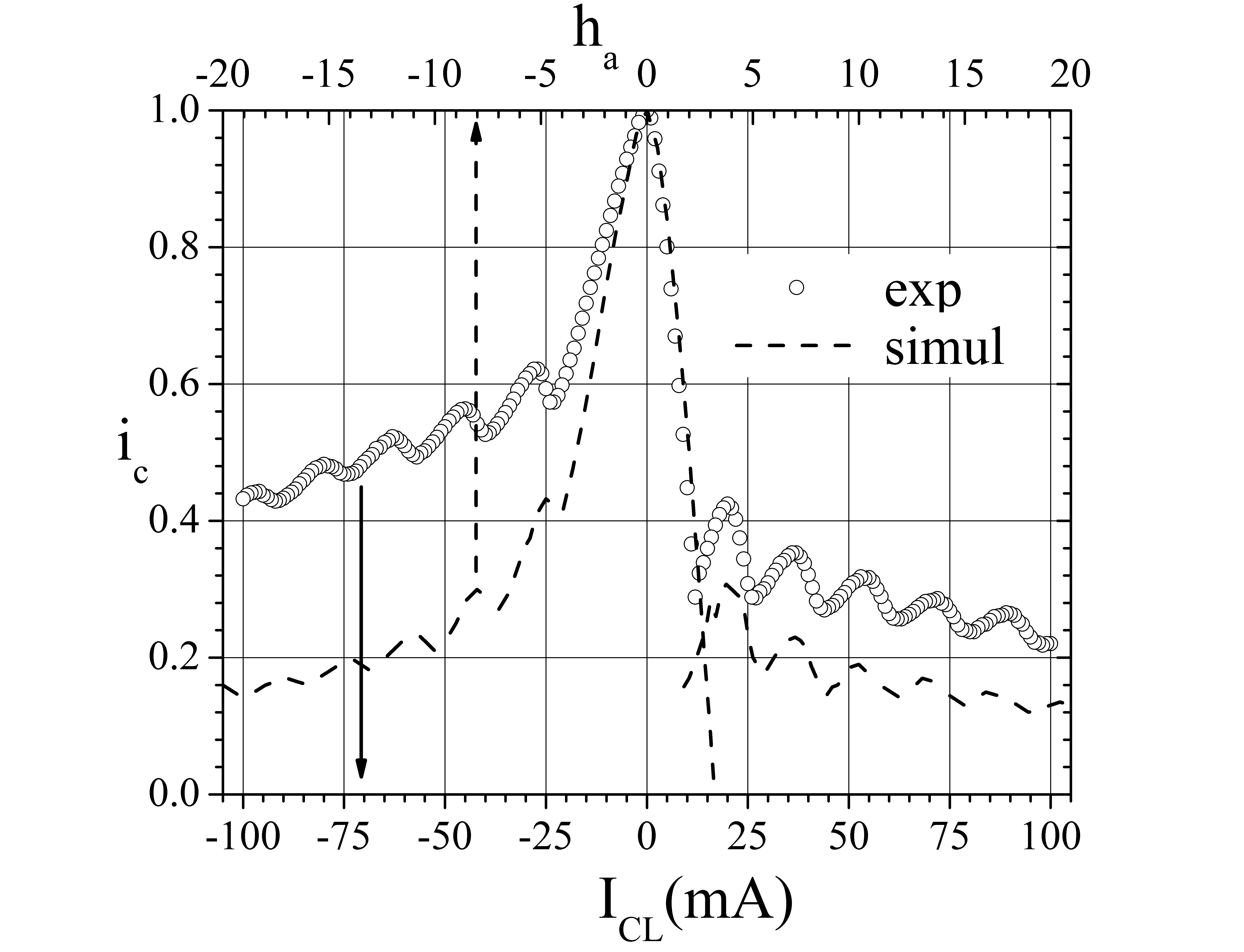}}
\caption{(a) Normalized MDP obtained with the control lines in the asymmetric mode (closed circles and lower horizontal scale) and the numerically computed MDP $i_c(h_{a})$ for an overlap junction with normalized length $l=4$ in a zero-mean magnetic field profile  $h(x)=2h_a x/l$ (dashed line and upper horizontal scale). (b) Normalized MDP obtained by feeding just one control line (closed circles and lower horizontal scale) and the numerically computed MDP $i_c(h_{a})$ for an overlap junction with $l=4$ in a field profile: $h(x)=h_a (x+l/2)/l$ (dashed line and upper horizontal scale).}
\end{figure}

\noindent The normalized MDP of the same sample in presence of the asymmetric field profile produced by the control line operated in the asymmetric mode is reported in Fig.3 (closed squares and lower horizontal scale) versus the common control line current amplitudes, $I_{CL}=I_{RCL}=-I_{LCL}$. We observe a pronounced  skewness despite the fact that the maximum $I_c$ value occurs for $I_{CL}=0$; the negative pattern (not shown here) reveals perfect symmetry with respect to the simultaneous inversion of both the junction and control lines currents. In the same plot we superimpose the numerically computed $i_c(h_a)$ dependence (open circles and upper horizontal scale) for an uniformly biased overlap junction with $l=4$ embedded in a linear zero-mean magnetic profile $h(x)=2h_a x/l$, so that $h(\pm l/2)=\pm h_a$, as reported in Ref.3. We observe agreement only at a qualitative level. We believe that the discrepancy can be ascribed to the fact that in our experiments the field gradient is unevenly distributed over the junction length, being rather 'squeezed' near its center. However, in the well-known modeling of a one-dimensional JTJ\cite{ferrel} the externally applied magnetic field only enters as the boundary conditions of a static sine-Gordon equation regardless of the particular field profile\cite{owen}; our results indicate that this approach does not capture all important physical details and should be used only for very long junctions when $l>>1$.

\noindent To further investigate the effect of non conventional magnetic profiles, in Fig.3(b) we show the MDP  resulting by biasing only one of the two control lines (closed squares and lower horizontal scale). The specular pattern (not shown) obtained by using the other control line indicates that this process is equivalent to a field inversion. Again we compare the experimental data with the numerically computed $i_c(h_a)$ dependence (open circles and upper horizontal scale) for an overlap junction with $l=4$ embedded in a uniform magnetic gradient: $h(x)=h_a (x+l/2)/l$, so that $h(-l/2)=0$ and $h(l/2)=h_a$. As expected for the reasons above, we get just  qualitative agreement.

\begin{figure}[ht]
\subfigure[ ]{\includegraphics[width=7cm]{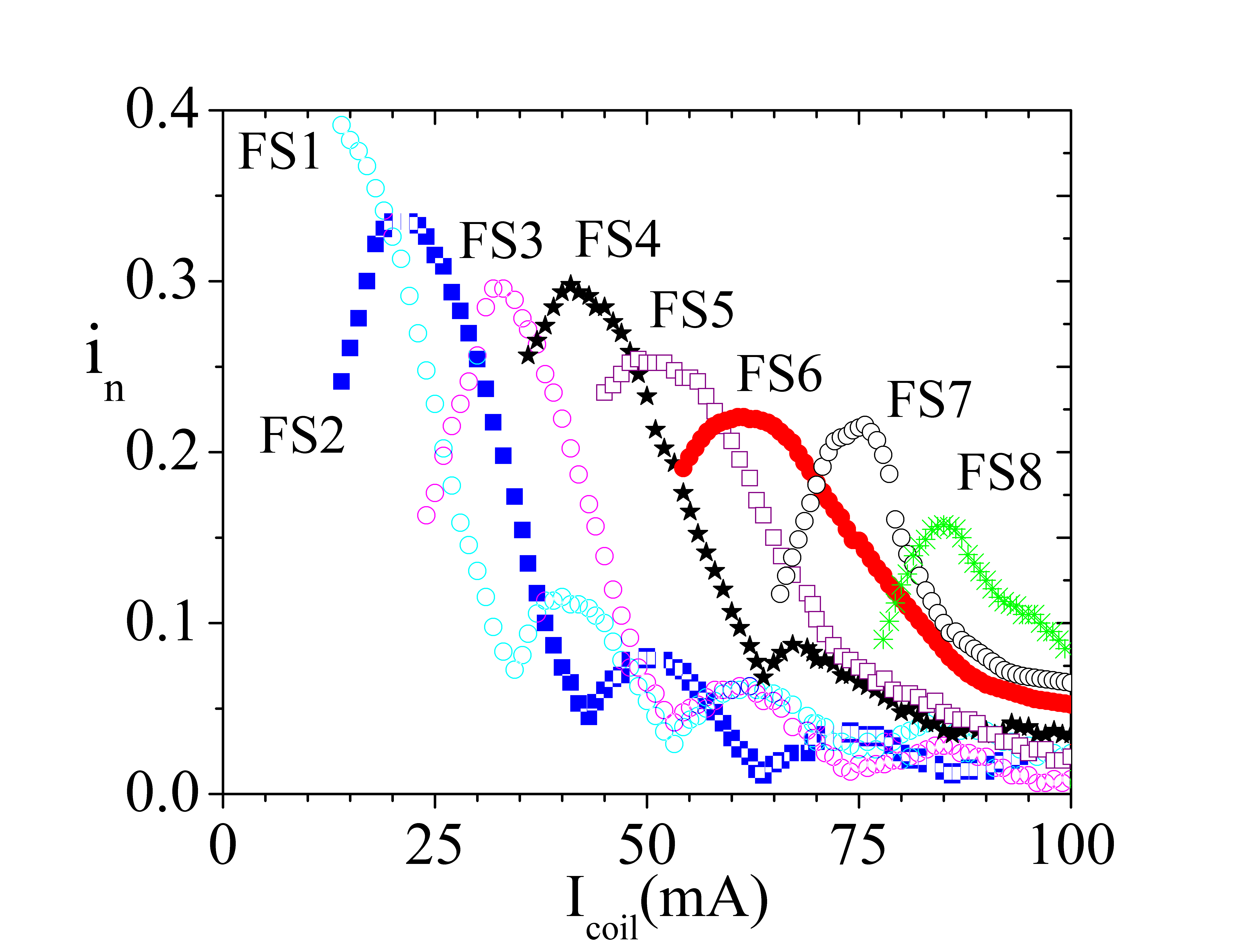}}
\subfigure[ ]{\includegraphics[width=7cm]{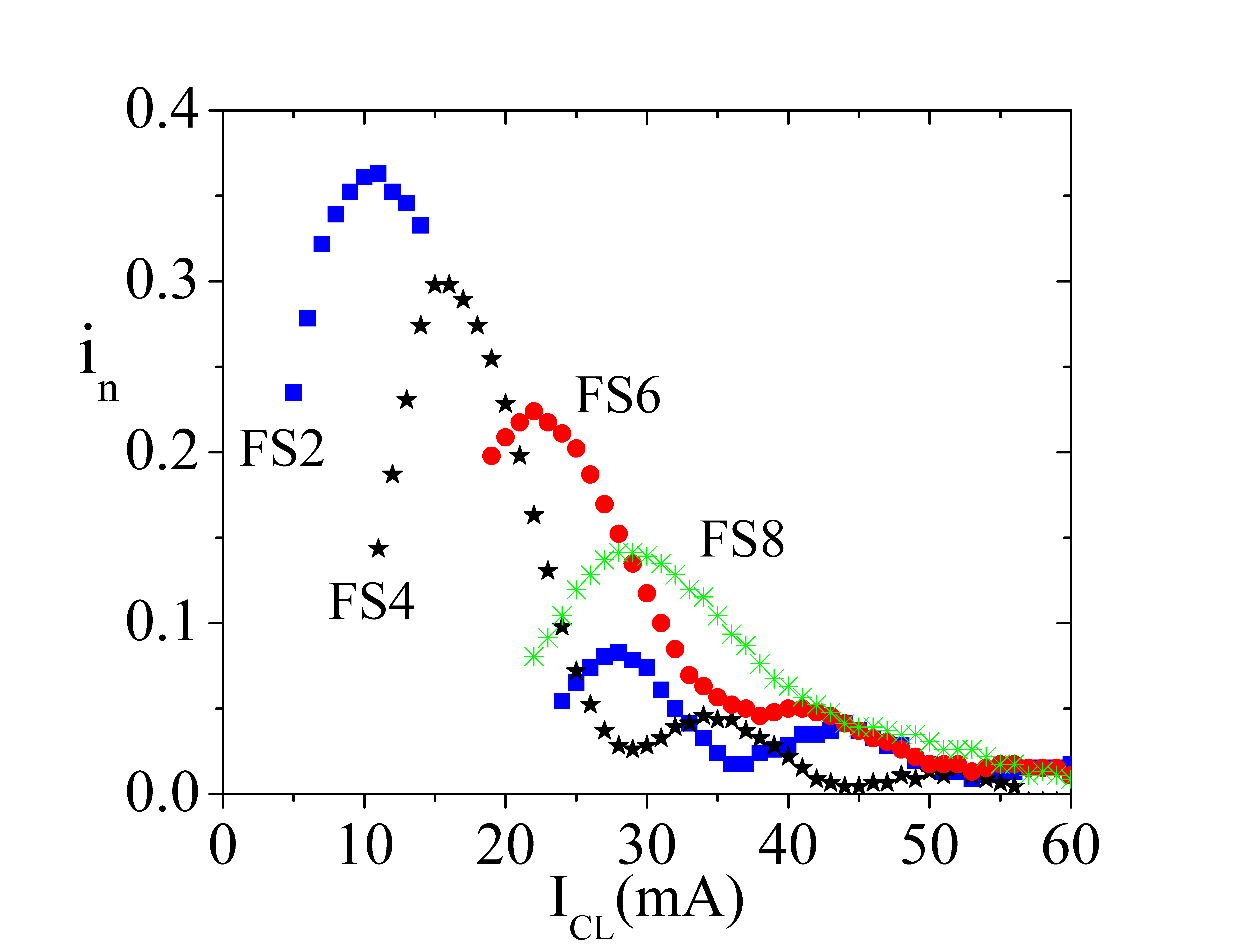}}
\caption{(Color online) Magnetic field dependence of the amplitudes of Fiske steps (a) in a uniform field and (b) in the inhomogeneous field generated by the control lines operated in the asymmetric mode. $T\approx 6 K$ and $I_c(0)=2.3 mA$.}
\end{figure}

\noindent A inhomogeneous magnetic field drastically modifies also the dynamics of a planar JTJ. Since the large idle region prevents the zero-field resonant fluxon motion\cite{ustinov}, we focused our attention on the magnetic resonant modes\cite{kulik} which manifest in the junction current-voltage characteristic as current singularities called Fiske steps\cite{fiske}. Figs.4(a) and (b) display the magnetic field dependence of the Fiske steps amplitudes for our sample in presence, respectively, of a uniform in-plane field and of the non-uniform field profile obtained in the asymmetric mode. These measurements were taken a $T\approx 6 K$ [$I_c(0)=2.3 mA$] since at lower temperature it was difficult to latch on the low order Fiske steps. We observe that in presence of an asymmetric profile only the even resonances survive; the full suppression of the odd steps is consistent with recent analytical calculations\cite{ciro} aimed to extend the Kulik theory\cite{kulik} for small junctions to a  magnetic field distribution of type $h(x)=h_a 2x/l$. We believe that the key ingredient of the odd step suppression is the asymmetry of the field profile. Surprisingly, {\it all} the resonances disappear in presence of the field profile proper of just one control line; indeed, Raissi {\it et al.}\cite{raissi} proposed that the reversing magnetic field sets extra conditions at the center of long JTJs which makes the creation of any standing electromagnetic waves impossible. Definitely, further investigation is needed to find a proper theoretical interpretation of these facts. 

\noindent To summarize, we have experimentally shown how the static and dynamic properties of Josephson tunnel junctions change when one abandons the standard assumption of a homogeneous magnetic field. New measurements with samples having both shorter and longer normalized length have already been planned. In the near future, the tuning of a shuttling fluxon by means of a magnetic field gradient will also deserve our attention.

\end{document}